# A point cloud processing method of mmWave radar over automotive scenario

Qingmian Wan, Hongli Peng, Xing Liao, Kuayue Liu

## I. Abstract

This paper introduces in detail the effective method of comprehensive target judgment by using radar RA map and point cloud map. Different output of radar can effectively judge the road boundary of target and the relative coordinates of target, avoid the error of output caused by excessive processing information, and greatly improve the processing efficiency of DBSCAN of the measured target.

## II. Introduction

Recently, millimeter-wave (mmWave) radar with lower price and better performance has been widely applied to mobile robots such as the wheeled robot, unmanned aerial vehicle (UAV), and unmanned surface vehicle (USV) [1]. Typically, automotive radar plays an important role in advanced driver-assistance (ADAS) systems for adaptive cruise control, collision avoidance, blind spot detection, lane change assistance, parking assistance, etc [2]. mmWave radar has the potential to replace LiDAR in robotic applications with the robustness to different lighting and weather conditions. Usually, the signal processing techniques of traditional radar detects the signal peaks from the radar echo data, decomposing the radar images into sparse point clouds. Since the point cloud obtained by mmWave radar is sparse, noisy and lacks intuitive information, point cloud processing such as segmentations and recognitions are researched to extract the target characteristics. Target recognition and classification of mmWave can be achieved using deep learning method. PointNets is used in literature [3] for bounding box estimation of the vehicles by distinguishing signals from clutter and vehicles. However, the direct usage of PointNets is not a good choice cause the design intention of PointNets is to learn the 3D spatial features of objects in LiDAR detection but the point cloud using mmWave radar detection is usually sparser with less information.

In this paper, a self-developed high-precision 4D mmWave radar system is introduced firstly. Based on the requirements of point cloud resolution and imaging quality, we have finished the overall design and optimized the system by analyzing the real-time detection performance, which enables the mmWave radar to achieve certain detection range and high-quality point clouds acquirement. Secondly, we propose a modified dynamic DBSCAN method to cluster the processed point cloud of targets, using the radar emitting equation and the target distribution features, which achieves better separation and recognition performance than the original point cloud. Finally, a neural network (NN) structure based on PointNet, using the range-azimuth map (RAM) for target segmentation and spatial coordinate restoration, is designed meticulously with the created dataset consisting of different testing scenes. Additionally, the trained model is evaluated on the collected classification dataset, achieving good performance in different complex scenarios. Then, we analyzed the performance of the model, indicating that it can decompose different categories of targets. We compared the radar resolution models of different categories and combined the different characteristics of point clouds and projected target maps. The major contributions of this paper are summarized as follows:

1. We designed a high-precision 4D radar system, generated high-quality RAM and optimized the imaging mode to obtain high precise 3D echo point clouds.

2. We recognized and removed the curb from sectorial image of mmWave radar in polar coordinate, which is transformed into the projection of Cartesian coordinate for the subsequent detection using semantic NN.

3. Multi-filter method for clustering and multi-frame overlapped technology for detection are proposed to increase the system performance.

4. A high-precision point cloud dataset of 4D mmWave radar is generated with the target information such as length, width, velocity, types and so on, which can be used for further research work in target recognition using mmWave radar.

III. RELATED WORK

IV. High quality mmWave radar point cloud generation

The mmWave radar point cloud imagery diagram is shown in Figure 1 which is proposed in our preliminary work]. As shown in Figure 1(a), the recorded echo data matrix from mmWave radar system is distributed in 3 dimensions which are range, channel and chirp dimension respectively. After the 2D FFT is performed in range-Doppler domain, several range-Doppler maps (RDMs) are obtained. Then constant false alarm rate (CFAR) detection, as shown in Figure 1(b), is used to detect the peak signal as the outputted target point, and another 2D FFT is performed to estimate the azimuth and elevation angles of the point target. The generated point cloud with parameters of range, Doppler, azimuth angle and elevation angle, as shown in Figure 1(c) and Figure 1(d), is the target geometry in polar coordinate and can be easily transformed into Cartesian coordinate as the final point cloud formation which is consistent with LiDAR point cloud, as shown in Figure 1(e) and Figure 1(f).

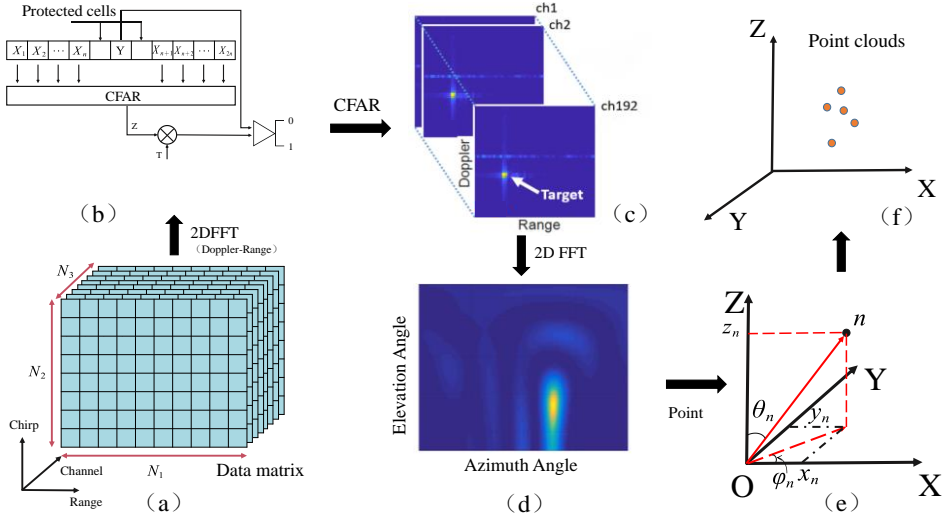

Fig. 1. The mmWave radar point cloud imagery diagram.

## A. RDM AND RAM GENERATIONS

Without loss of generality, the principle of millimeter wave radar point cloud imagery is briefly introduced in this section below and the detailed description can be found in our preliminary work. Normally, the 4D mmWave Radar emits frequency modulated continuous wave (FMCW) signal in time division multiplex (TDM) mode. The total echo data matrix from the scenario after differential frequency reception between the $m$-th transmitter and the $n$-th receiver can be written as:

$$s_{mn}(t, t_a) = \sum_{q=1}^{Q} \sigma_q \cdot \exp\left(-j\frac{2\pi R_{mn,q}(t_a)}{\lambda}\right) \cdot \exp\left(-j2\pi k_r \left(\frac{R_{mn,q}(t_a)}{c}\right) t\right) \\ \cdot \exp\left(j\pi k_r \left(\frac{R_{mn,q}(t_a)}{c}\right)^2\right) \quad (1)$$

where $t$ and $t_a$ are slope time and azimuth time respectively. $Q$ is the number of scatterers according to the scattering center hypothesis. $f_c$ is the carrier frequency, $k_r$ is the chip rate of the transmitted signal. $c$ is the light velocity. $R_{mn,q}(t_a)$ is the sum of the distances from the $q$-th scatterer to the $m$-th transmitter and the $n$-th receiver respectively.

Assuming $v_{r,q}$ is the radial velocity between the $q$-th scatterer and the radar, the RDM of each channel can be obtained through 2D FFT in range-Doppler domain, which can be written as:

$$S_{mn}(f, f_d) = T_a \cdot T_p$$

$$\cdot \sum_{q=1}^{Q} \sigma_q \cdot \exp\left(-j2\pi \frac{R_{mn,q}(t_{q_0})}{\lambda}\right) \cdot \text{sinc}[T_a(f_d - f_q)] \quad (2)$$

$$\cdot \text{sinc}\left[T_p\left(f - 2k_r \frac{R_{mn,q}(t_a)}{c}\right)\right]$$

where $f$ is the modulated frequency of the baseband signal. $f_q$ is the Doppler frequency caused by the target movement and $f_q = 2v_r/\lambda$. $T_a$ is the coherent processing time during the target movement and $T_p$ is the chirp signal pulse duration. $\lambda$ is the wave length. $R_{mn,q}(t_a)$ is the radial movement between the $q$-th scatterer and the radar with $R_{mn,q}(t_a) = R_{mn,q}(t_{q_0}) - v_{r,q}t_a$.

The RDM $\bar{s}_{mn}(f, f_d)$ in Equation (2) is always used in automotive 4D mmWave radar as the dimensions of range and velocity which can be direct calculated from $f_q = 2v_r/\lambda$. Another two dimensions are the estimated angles of the point target in azimuth and elevation directions respectively.

RDM is the foundation of the 4D mmWave radar imagery where the point target information can be detected and extracted, commonly used in the automotive areas. However, the stationary objects are neglected since the target moving assumption which can be represented in the RDMs. The stationary objects without movement are distributed in the zero Doppler cell of the RDM which can hardly be distinguished.

From Equation (1), we ignore the high-frequency terms which can be neglected in non-focused imaging scenarios and derive the echo signal formation of distributed target as:

$$s(r_{m,n}) = \sum_{a=1}^{A} \sigma_a \cdot \exp\left(-j\frac{2\pi r_{m,n}}{\lambda}\right) \approx \sum_{a=1}^{A} \sigma_a \cdot \exp\left(-j\frac{4\pi r}{\lambda}\right) \quad (3)$$

where $r_{m,n}$ is the sum of the distances from the $a$-th element in surveillance scenario to the $m$-th transmitter and the $n$-th receiver respectively with stationary assumption, and $r \approx r_{m,n}/2$ with the phase center approximation (PCA) algorithm. $A$ is the total number of the pixel elements in the detection area.

By translating Equation (3) from the polar coordinate to the Cartesian coordinate, we can get:

$$s(k_x, k_y) = \sum_{x=1}^{A_x} \sum_{y=1}^{A_y} \sigma(x, y) \cdot exp(-jk_x \cdot x) exp(-jk_y \cdot y) \quad (4)$$

where $k_x = 4\pi \cdot \cos\theta /\lambda$ and $k_y = 4\pi \cdot \sin\theta /\lambda$.

When the virtual array distribution which can be calculated form the PCA algorithm is uniform, the $\sigma(x,y)$ can be obtained from the 2D Fourier transformation of $s(k_x, k_y)$ using the small angle approximation. $\sigma(x,y)$ is distributed in Cartesian coordinate and can be easily translated into polar coordinate as $\sigma(r,\theta)$ which is called RAM. Compared to the RDM, RAM has better ability to represent the stationary characteristics of the detected environment.

B. POINT CLOUD GENERATION AND OPTIMIZATION

The key point to generate accurate point cloud of target using 4D mmWave radar is the design of antenna array to distinguish target points in azimuth and elevation directions, with high resolution RDMs containing range and velocity information. With the incensement of the number of the MIMO transceiver antennas, the virtual antenna elements are increased significantly. The desired function of the 4D mmWave radar is then upgraded and promised to better reconstruction of the target details including distributions and contours, while the conventional radar still follows the algorithm of target detection.

Based on the principle of minimum redundancy array (MRA), the virtual array formed by 12 TXs and 16 RXs is depicted in Figure 2(a) and the ambiguity function map (AFM) is analyzed and shown in Figure 2(b) which is computed from simulation method.

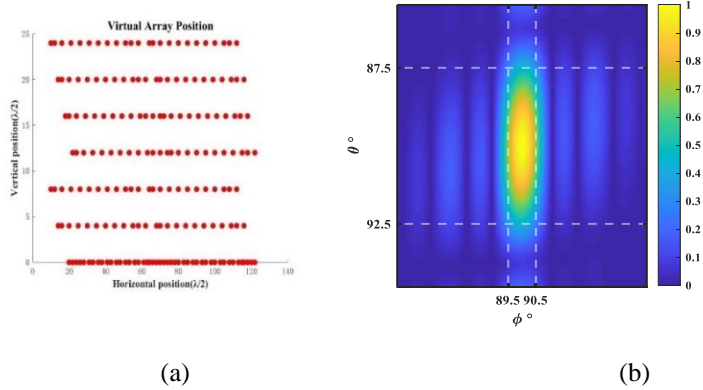

(a)                                                (b)

Fig. 2. The resolution analysis of the designed array with (a) the virtual array and (b) the ambiguity function map.

The designed angular resolutions using this antenna array achieve 1 degree in azimuth and 5 degrees in elevation respectively. This design provides the mmWave radar sufficient ability to distinguish targets in azimuth and elevation direction.

Theoretically, the signal of weak target can be detected by the system through the low noise amplifier (LNA). However, the power of detected signal must be larger than the noise. Considering the system loss, the valid detection range can be written as:

$$R_{max} = \left[\frac{P_t G^2 \lambda^2 \sigma}{(4\pi)^3 K B_0 T_0 F_0 \text{SNR}_{min} L}\right]^{1/4} \quad (5)$$

where $KB_0 T_0 F_0 \text{SNR}_{min} L$

Therefore, the detected target points satisfy two conditions:

(1) The echo power of the target is larger than the noise base.

(2) The target point presents a peak form in RDM.

Only the echo power of the target exceeds the CFAR threshold and the system noise base can the valid detected. Therefore, the further points detected are sparser with less details.

As can be seen from the range profile which is presented in Figure 3, the reflected intensity of the objects which have same radar cross section (RCS) exponentially decreases with the increase of the distance.

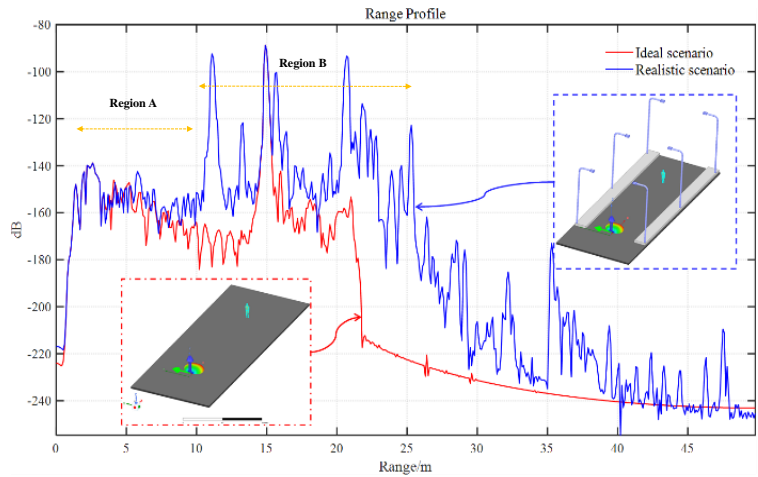

Fig. 3. Range profile comparison of ideal and realistic scenarios.

With the distance increase, the reflection intensity of radar point cloud will decrease significantly. The boundary points can hardly be detected when using CFAR detector with the fixed threshold. Then the density of the point cloud is decreased with the distance increase, which is shown in Figure 4.

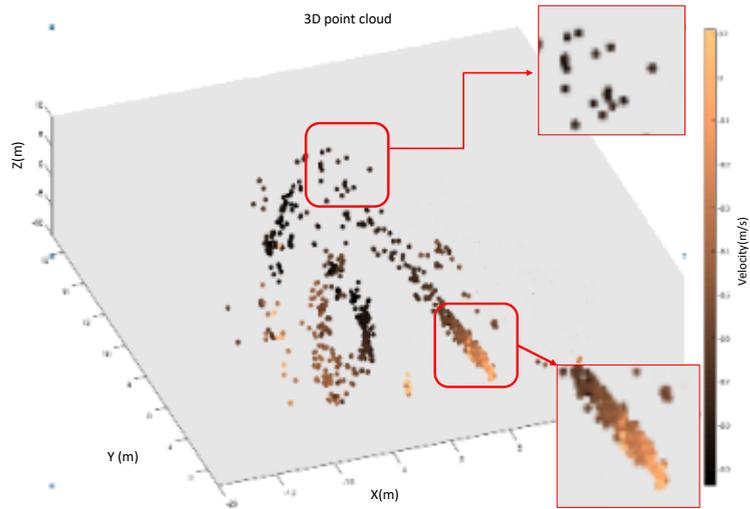

Fig. 4. The density of the point cloud at different distances.

Therefore, vehicles and roadside points will become confused, resulting in detected size errors in areas with high point cloud density as shown in the figure 5. According to the radar equation, the reflection power decreases exponentially to the fourth power with increasing distance. Therefore, the same target will have different point cloud reflection power at different distances. The categories of point cloud become more difficult to be distinguished as the distance increases, and the echo signal approaches the system noise base, leading to loss of details. As shown in Figure 5, the roadside point cloud is confused with the vehicle target point cloud.

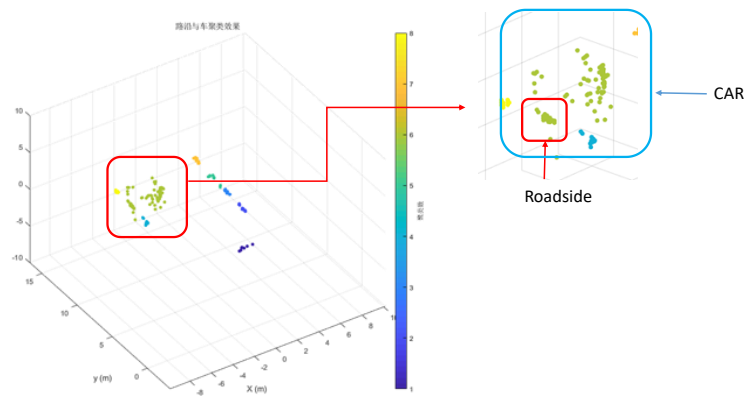

Fig. 5. The extracted points of a car in density point cloud distance.

Since the RAM is a whole projection of detected scene in 1D plan, the demand for point cloud density is reduced, which is shown in Figure 6. The 1D projection can completely represent all the targets in surveillance area. Therefore, it can be considered to first extract the road edges in RAM. Additionally, a unified point cloud clustering method is also considered for the same type of targets at different distances.

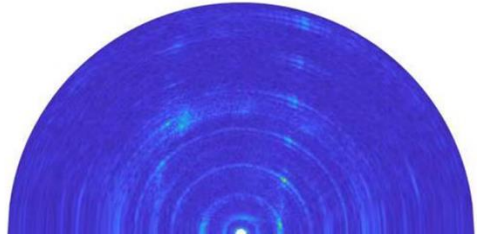

Fig. 6. The RAM containing roadside in polar coordinate.

As mentioned above, the same RCS target reflection at different distances will have different point densities. Then a dynamic DBSCAN clustering method is proposed whose diagram is described as:

(1) Calculate the minimum radius value and minimum cluster number using the k-distance method.

(2) Perform DBSCAN clustering on the point cloud according to the velocity differences.

(3) Perform DBSCAN clustering on point clouds of each velocity cluster generated from step (2) in Cartesian coordinate.

(4) Remove clusters with small number of points and unreasonable shapes.

The minimum radius established depends on the a $k$-distance graph with $k$ selected according to the number of data matrix dimensions, for example, twice of the number of data matrix dimensions. The $k$-distance graph is a topology diagram of the distance from each point to the $k$-th point closest. After sorting these distances, the minimum radius is the distance value of the inflection point which can be used in the clustering as the radius threshold. A typical example embodiment is illustrated in Figure 7.

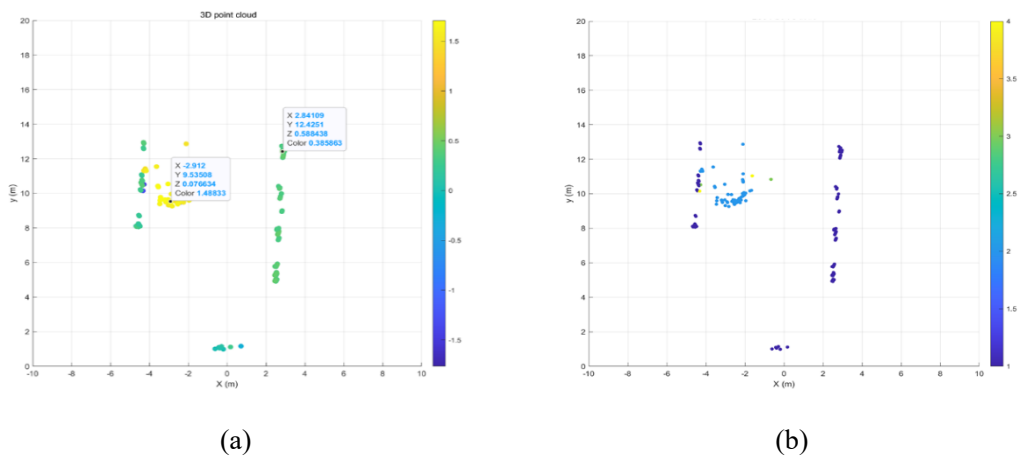

(a)          (b)

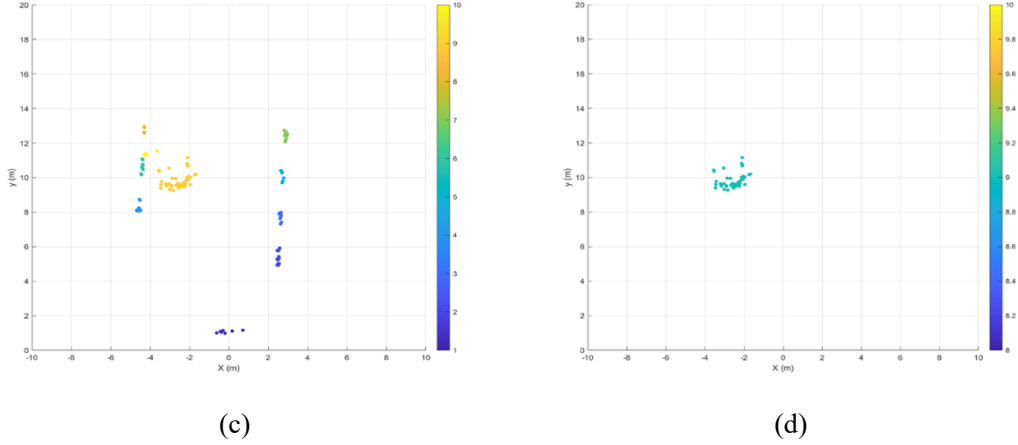

(c)                                          (d)

Fig. 7. The point cloud of (a) original generated, (b) after velocity clustering, (c) after the second clustering and (d) the final produced.

## V. Target information extraction method

Based on the output characteristics of mmWave radar targets, in order to obtain better classification results for complex scenarios, different types of output data are combined using the UNET model to identify the roadside in the bird-eye view with RAM as input images. Combining the advantages of spatial-stereo radar point cloud points, the output data of different models is separated and used. The final output is fused into the Cartesian coordinate, using another designed NN, with information such as categories, sizes, locations, etc.

### A. IMAGE SEGMENTATION AND FILTERING

In order to classify the detected targets, the RAM and the point cloud generated from RDMs of the mmWave radar are transferred into a segmentation module, followed by point cloud cluster filtering. Extracting spatiotemporal information from 4D mmWave radar into point clouds, the proposed method is illustrated in Figure 8. The proposed method has 2 major parts of a pixel feature learning UNET module to extract pixel information from RA images and a spatial fusion module with different coordinate systems.

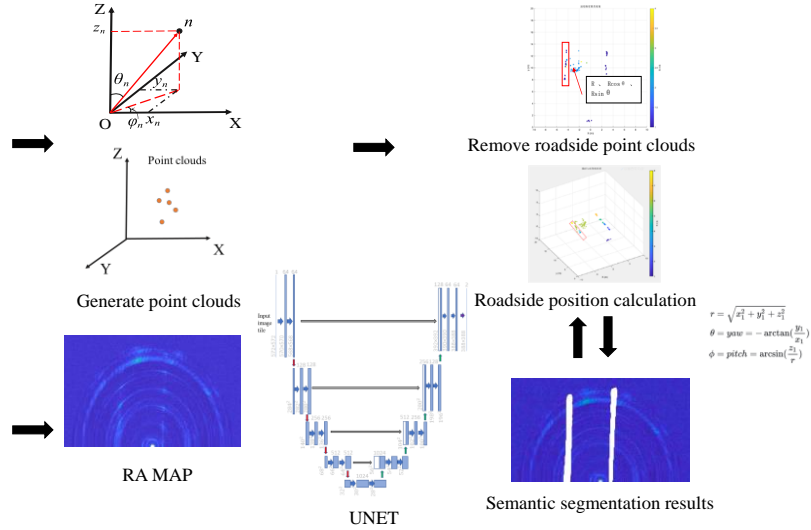

Fig. 8. The proposed segmentation and filtering method.

Since the RAM is still quite noticeable which can be seen from Figure 8, the target features are preserved with the distance extending. RAM needs only 2D information of the target, so the target is clearly represented in the RAM. The RAM detection network structure is shown in Figure 9. It consists of a contraction path (left) and an expansion path (right). The contraction path follows the typical architecture of convolutional networks. It includes repeated applications of two 3x3 convolutions (unfilled convolutions), with each convolution followed by a rectified linear unit (ReLU), as well as a 2x2 max pooling operation with a step size of 2 to perform the down-sampling operation. In each down-sampling step, we will double the number of feature channels. Each step in the extension path involves up-sampling of the feature map, followed by a 2x2 convolution, halving the number of feature channels. After cutting the number of feature channels by half using a 2x2 convolution, the feature channels are combined with the corresponding cropped feature maps and two 3x3 convolutions in the contraction path, with each convolution result is activated using a ReLU. Due to the loss of boundary pixels in each convolution, the cropping operation is necessary to be taken. In the final layer, a 1x1 convolution is used to map each 64 component feature vector into the desired categories. There are a total of 23 convolutional layers in the network. It is crucial to choose the size of the input tile in order to achieve seamless tiling of the output segmentation image. It is important to choose the size of the input tiles so that all 2x2 max pooling operations are applied to even numbered layers, which have the same size in two dimensions of the images.

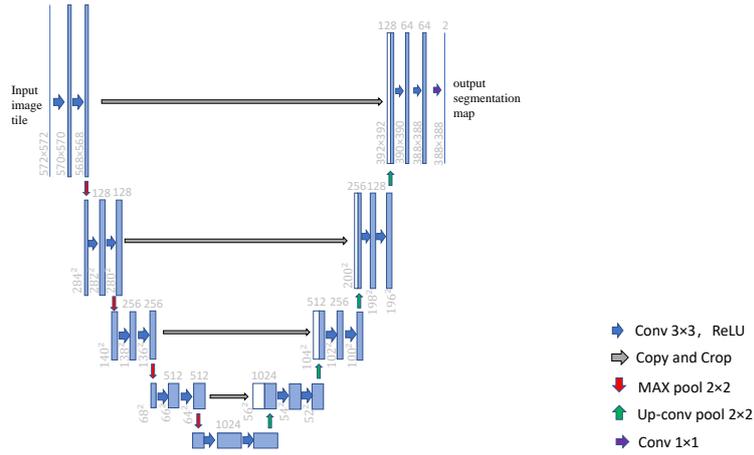

Fig. 8. The RAM detection network structure.

Additionally, multiple frame overlay is adopted to provide results with higher confidence, which can increase the point cloud accuracy and reduce the noise level. The point cloud of each frame of the mmWave radar contains individual points such as clutter points and target points. The normalization is necessary to transform the heterogeneous data into a same shape. The input data mainly consists of 4 dimensional information, which are 3D spatial information and 1D velocity information. Coordinate transformation is carried out in the subsequent target space coordinates. The velocity is involved in the clustering operation. Then the target trajectory can be determined using any tracking method.

B. POINT CLOUD INFORMATION EXTRACTION

The point cloud with Doppler velocity has 4D information to be processed. Firstly, the point cloud generated after dynamic DBSCAN is sampled using farthest point sampling (FPS) algorithm. Then the ball query is taken out which samples some points in a fixed radius distance around the query point and outputs the sampling indices. According to the sampling indices, the sampling points in the first step can be selected. Meanwhile, a number of points around the selected point are also selected and merged into the point set as the grouping result. Finally, the updated points are outputted with the merged point set. Then, a Bi-LSTM network is introduced as the sequential information learning module with the attention weight features, and a 4D-PointNet inspired by PointNet is also designed to predict the point features. For the 4D mmWave radar point cloud, different feature alignment is performed with feature transformation matrix using a mini network called T-net. The detailed network structure is illustrated in Figure 9.

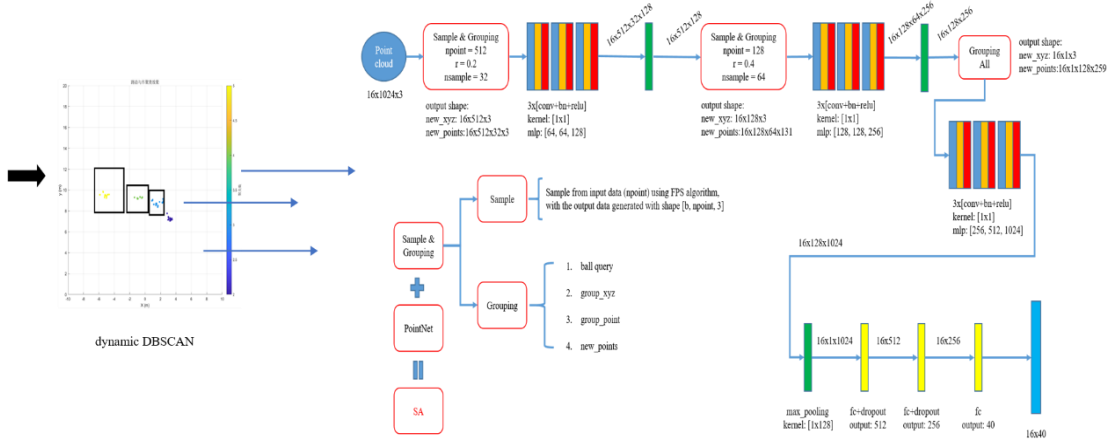

Fig. 8. The network structure of point cloud information extraction.

C. LOSS FUNCTIONS

In the context of our image segmentation task, we have designed a composite loss function that combines both the Dice Loss and the Focal Loss. This approach is commonly employed in image segmentation to enhance segmentation performance and effectively address the issue of class imbalance. The Dice Loss, a well-established metric for measuring the degree of overlap between predicted and ground truth segmentation masks, encourages the model to generate accurate segmentations. It is formulated as follows:

$$\text{Dice Loss} = 1 - \frac{2 \times \sum_i (\text{Prediction}_i \times \text{Ground Truth}_i)}{\sum_i \text{Prediction}_i^2 + \sum_i \text{Ground Truth}_i^2} \quad (6)$$

The Dice Loss focuses on maximizing the overlap between predicted and actual segmentations, promoting fine-grained segmentation accuracy.

In addition to the Dice Loss, we have incorporated the Focal Loss to mitigate the challenges posed by class imbalance. The Focal Loss introduces a dynamic scaling factor that emphasizes hard-to-classify samples, thereby improving the model's sensitivity to rare or critical classes. The Focal Loss can be expressed as:

$$\text{Focal Loss} = -\alpha \times (1 - \text{Prediction})^\gamma \times \log(\text{Prediction}) \quad (7)$$

where $\alpha$ is the factor which controls the balance between positive and negative class samples. $\gamma$ is the factor which adjusts the focusing factor to give more weight to hard samples.

Moreover, we have introduced the flexibility of adjusting the relative importance of different classes within the loss function by utilizing class weights This feature allows us to assign varying weights to individual classes, ensuring that the model's training process is attentive to the specific requirements and imbalances within our segmentation problem.

By integrating these components, our composite loss function encourages precise and robust image segmentation, even when faced with challenging class distributions, ultimately enhancing the model's performance on our segmentation task.

In this paper, we employ the Cross-Entropy loss function as the training objective for our deep learning model. Cross-Entropy loss is widely used in multiclass classification problems and plays a pivotal role in our research. This loss function helps quantify the dissimilarity between our model's outputs and the actual labels.

The mathematical expression of the Cross-Entropy loss is as follows:

$$L(y,p) = -\sum_{i=1}^{n}(y_i \log(p_i) + (1-y_i)\log(1-p_i)) \qquad (8)$$

where $L(y,p)$ represents the Cross-Entropy loss, $n$ denotes the number of categories in the classification task, $y_i$ stands for the true binary labels, and $p_i$ represents the predicted probabilities for the corresponding categories by the model.

Our choice is based on the versatility and effectiveness of the Cross-Entropy loss function, particularly in multiclass classification problems, guiding our model to learn in the right direction. By employing the Cross-Entropy loss function, we can better train our model, allowing it to classify more accurately, thus establishing a solid foundation for our research.

## VI. EXPERIMENT AND EVALUATION

To verify the proposed method in this paper, we carried out experiments with the prototype mmWave system which is shown in Figure 9(a). The deployed scenario and system are shown in Figure 9(b) and some test scenarios containing targets are shown in Figure 9(c).

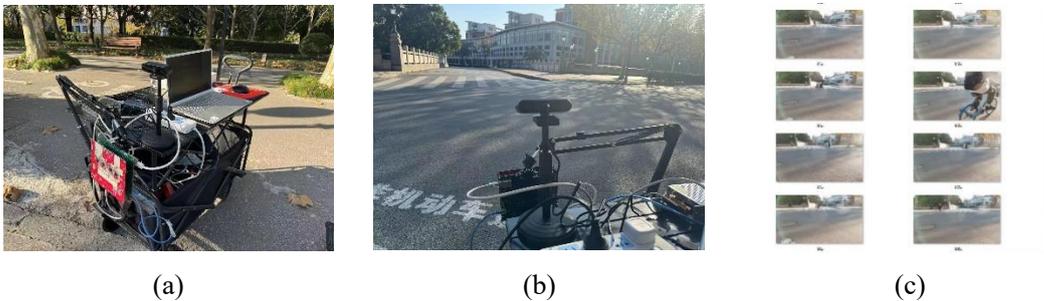

(a)          (b)          (c)

Fig. 9. (a) The prototype mmWave system, (b) the deployed scenario and system and (c) some test scenarios containing targets.

### A. IOU TESTING

The experimental data contains objects of the roadside and the target points in the RAM and the generated point cloud respectively, which are transferred into the same coordinate system. The transformation accuracy is tested by introducing the criterion of intersection over union (IoU) which is described as $(b_1 \cap b_2)/(b_1 \cup b_2)$ with two areas of $b_1$ and $b_2$.

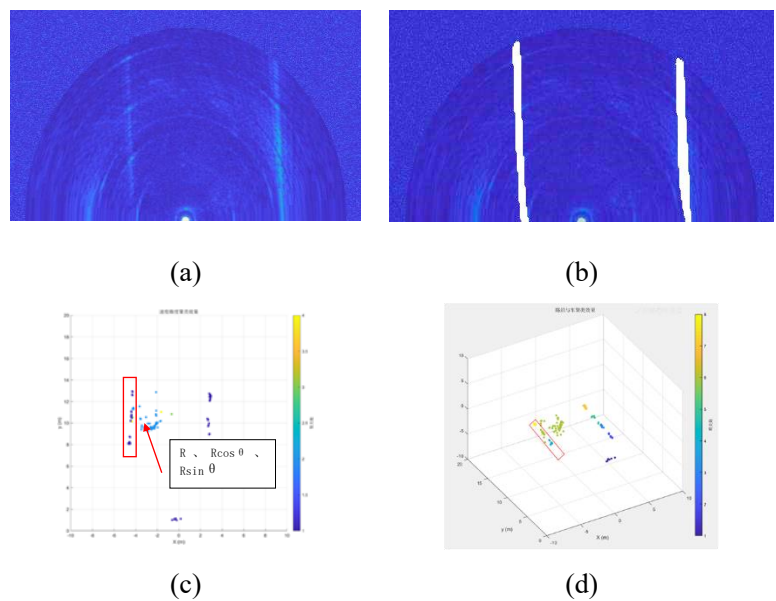

Fig. 10. (a) The original RAM, (b) the RAM after roadside segmentation using UNET, (c) the 2D point cloud contains roadside and (d) the 3D point cloud contains roadside.

The test IoU rate over the roadside area is shown in Table 1, where we can get the IoU rate is larger than 90%. That means the segmentation accuracy of the road edges is high. The road edge extracted from RAM can overlap well with the point cloud map. The IoU rate will increase when the fixed threshold is relaxed.

Tab. 1. The tested IoU rate over the roadside area.

| area 1 | area 2 | | IoU rate |
|---|---|---|---|
| left roadside area | X (m) | Y (m) | |
| left top | -6.829 | 0 | |
| right top | -5.566 | 0 | >90% |
| left bottom | -5.7832 | 12.36328 | |
| right bottom | -4.8427 | 12.1875 | |

B. POSITIONING ACCURACY

To analyze the target positioning accuracy, a test scenario is selected as shown in Figure 11. A corner reflector is placed between two stationary cars, and the distances from the corner reflector to the left vehicle and the right vehicle are measured and recorded.

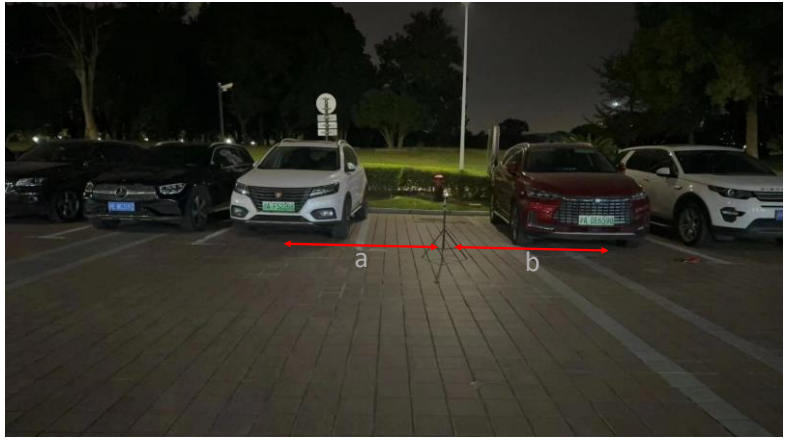

Fig. 11. The test scenario of target positioning.

The positioning accuracy is analyzed in Table 2. It can be seen that the positioning error is distributed from 0 to a maximum value of about 15%. The overall error is within 30cm, reaching an acceptable value in automotive applications.

Tab. 2. Target positioning accuracy analysis.

| extracted data index | distance a (m) | distance b (m) | error a | error b |
|---|---|---|---|---|
| 1 | 2.56 | 2.325 | 12.53% | 4.03% |
| 2 | 2.275 | 2.015 | 0.00% | -9.84% |
| 3 | 2.285 | 1.885 | 0.44% | -15.66% |
| 4 | 2.22 | 2.065 | -2.42% | -7.61% |
| 5 | 2.52 | 2.38 | 10.77% | 6.49% |
| 6 | 2.2 | 2.15 | -3.30% | -3.80% |
| 7 | 2.5 | 2.095 | 9.89% | -6.26% |
| 8 | 2.55 | 2.055 | 12.09% | -8.05% |
| 9 | 2.31 | 2.22 | 1.54% | -0.67% |
| 10 | 2.245 | 2.315 | -1.32% | 3.58% |
| 11 | 2.565 | 2.245 | 12.75% | 0.45% |
| 12 | 2.5 | 2.345 | 9.89% | 4.92% |
| 13 | 2.26 | 2.4 | -0.66% | 7.38% |
| 14 | 1.975 | 2.435 | -13.19% | 8.95% |
| 15 | 1.925 | 2.425 | -15.38% | 8.50% |
| 16 | 2.13 | 2.395 | -6.37% | 7.16% |
| averaged value | 2.31375 | 2.234375 | | |
| measured value | 2.275 | 2.235 | | |

C. TARGET CLASSIFICATION ACCURACY

The target classification accuracy is tested under the scenario setup which is shown in Figure 12. In order to obtain a mmWave radar dataset suitable for semantic segmentation and target classification, the data collected 80 times containing different targets, such as pedestrians, bicycles and cars.

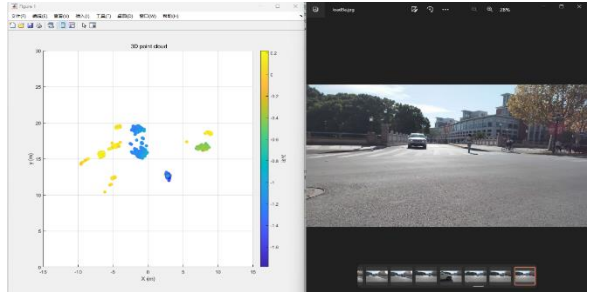

Fig. 12. The test scenario of target classification.

The collected data is divided into training dataset and testing dataset with the ratio of 3:1. After 1000 iterations of training, the tested accuracy over the testing dataset of target classification in complex scenarios achieves 93.10%. In comparison, it achieves 80.10% using RADNET and 60.78% using pointnet++. The confusion matrix of 4 types target classification is shown in Table 3.

Tab. 3. The confusion matrix of 4 types target classification.

|  | car | roadside | bicycle | pedestrian |
|---|---|---|---|---|
| **car** | 100% | 0 | 0 | 0 |
| **roadside** | 0 | 95% | 0 | 0 |
| **bicycle** | 10% | 0 | 80% | 10% |
| **pedestrian** | 0 | 0 | 0 | 100% |

The learning rate curves of the loss and accuracy are shown in Figure 13(a) and Figure 13(b).

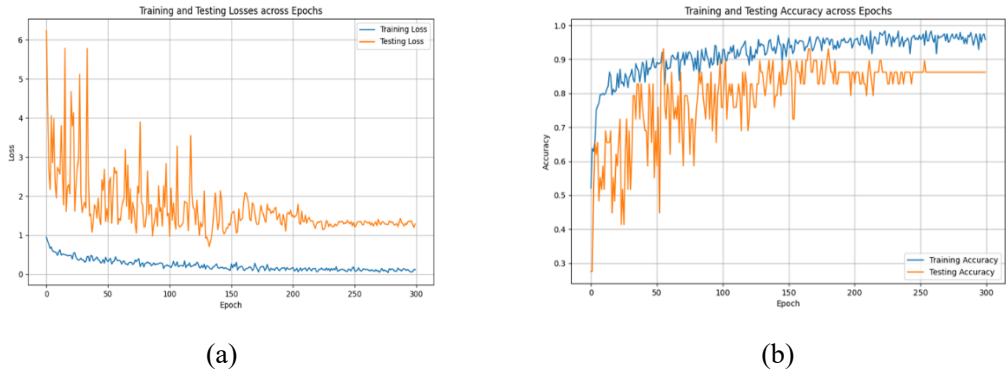

(a)                    (b)

Fig. 13. The learning rate curves of (a) the loss and (b) the accuracy.

## VII. CONCLUSION

In this paper, a point cloud processing method of mmWave radar over automotive scenario is proposed, containing the parts of high quality mmWave radar point cloud generation and target information extraction. The RAM is introduced in our work, which is different from the traditional mmWave radar system used in the automotive applications, to distinguish the stationary objects in the surveillance scenarios. The dynamic DBSCAN is adopted to promote the efficiency of the point target clustering. An elaborately designed network is proposed to extract the target information which shows better performance than RADNET and pointnet++ through the experiment analysis.

**QingMian Wan** received the B.S. degree in Optoelectronic Information Science from University of Shanghai for Science and Technology, Shanghai, China, and the M.S. degree in electrical engineering from Tongji University, Shanghai, China, in 2016. He works as a researcher in the Shanghai Automotive Industry Corporation Technology Center Autonomous Driving Department. He develops 4D millimeter-wave radar hardware and is responsible for autonomous driving algorithm optimization and environmental perception and other related research areas. He has authored or coauthored more than 10 articles in high-quality journals and international conferences.and all authors may include biographies. Biographies are often not included in conference-related papers.

**HongLi Peng** (M'10-SM'18) was born in 1966. He received the B.S., M.S. and Ph.D. degrees in electromagnetic field and microwave technique from Xidian University, Xi'an, China, in 1988, 1991, and 2005, respectively. From 1994 to 1999, he was a Senior Researcher, and the Project Leader with National Telemeter Center, Xi'an, China, for conformal antenna and its system designing. In June 1999, he joined the ZTE Corporation, Shanghai, China, as a Scientific Researcher and the Project Leader, where he contributed 11 essential patents. Since October 2008, he has been an Associate Professor in electromagnetic fields and microwave techniques with the School of Electronic Information and Electrical Engineering, Shanghai Jiao Tong University (SJTU). He has authored or coauthored more than 95 technical papers and owned over 42 patents. He published a book on MIMO indoor channel modeling. His current research interest mainly includes the tunable RF & microwave passive circuits research, re configurable compact antennas/array analysis and design, spatial wireless channel modeling.

Dr. Peng was a session co-chair of the 2011 IEEE Electrical Design of Advanced Packaging and Systems Symposium (EDAPS'2011), technically sponsored by IEEE CPMT Committee, a program committee co-chair of the 2012 IEEE International Conference on Cloud Computing and Intelligent Systems (2nd IEEE CCIS2012), a program committee co-chair of the 2012 IEEE International Conference on Cloud Computing and Intelligent Systems (2nd IEEE CCIS2012) and a workshop co-chair of 2020 Chinese Microwave Week Committee. He is a Reviewer of many international journals, including eight IEEE TRANSACTIONS and Letters. He earned the Standardization Award of China in 2005 and 2006, respectively. He received the Science and Technology Promotion Award of the first class from the local Shanghai Government of China in 2011, the Okawa Foundation Research Grant of Japan in 2018, and the Best Paper Awards, respectively, from the 2012 IEEE International Conference on Microwave and Millimeter Wave Circuits and System Technology and International Conference on Microwave and Millimeter Wave Technology 2019.

**Xing Liao** received the B.S. degree in electronic information science and technology from University of Electronic Science and Technology of China, Chengdu, China, in 2020, and the M.S. degree in electronic science and technology from Shanghai Jiao Tong University, Shanghai, China, in 2023. His current research interests include antenna and antenna array design, mmWave radar design.

**KuaYue Liu** received the B.Eng. degree in communication engineering from Tongji University, Shanghai, China, in 2021. Since 2021, he has been a Master student with the State Key Laboratory of Radio Frequency Heterogeneous Integration, Shanghai Jiao Tong University, Shanghai, China. His main fields of research interest include millimeter wave detection and perception and hardware design of 4D imaging millimeter wave radar.